\documentstyle[aps,graphicx,floats,prb]{revtex}

\def\u2{$\langle u^{2} \rangle$}

\def\qua3{{\tiny $\frac{3}{4}$}}

\def\iga{In$_x$Ga$_{1-x}$As}
\def\stz{ZnSe$_{1-x}$Te$_{x}$}
\def\as{a\sp{**}}
\def\sb#1{$_{#1}$}
\def\sp#1{$^{#1}$}

\begin{document}
\draft


\wideabs{

\title{Local Atomic Strain in \stz\  from High Real Space Resolution Neutron Pair Distribution Function Measurements}
\author {P. F. Peterson, Th. Proffen, I.-K. Jeong, S.~J.~L. Billinge}
\address{Department of Physics and Astronomy and Center for Fundamental Materials Research,\\
          Michigan State University, East Lansing, Michigan 48824-1116, USA}
\author{K.-S. Choi, M.~G. Kanatzidis}
\address{Department of Chemistry, Michigan State University\\
          East Lansing, Michigan 48824-1116, USA}
\author{P.~G. Radaelli} 
\address{ISIS, Rutherford Appleton Laboratory\\
          Chilton, Didcot OX11 0QX, UK}
\date{September 22, 2000} 

\maketitle


\begin{abstract}

High real-space resolution atomic pair distribution functions (PDFs)
have been obtained from \stz\ using neutron powder diffraction.
Distinct Zn-Se and Zn-Te nearest neighbor (nn) bonds, differing in
length by $\Delta r= 0.14$~\AA , are resolved in the measured PDF
allowing the evolution with composition of the individual bond-lengths
to be studied. The local bond-lengths change much more slowly with
doping than the average bond-length obtained crystallographically.
The nn bond-length distributions are constant with doping but
higher-neighbor pair distributions broaden significantly indicating
that most of the strain from the alloying is accommodated by
bond-bending forces in the alloy. The PDFs of alloys across the whole
doping range are well fit using a model based on the Kirkwood
potential.  The resulting PDFs give excellent agreement with the
measured PDFs over the entire alloy range with no adjustable
parameters.
\end{abstract}
} 


\section{Introduction}

Ternary alloys, such as \stz, are technologically useful because the
band-gap can be tuned between the end-member values as the
composition, $x$, is varied. This makes the proper characterization of
these materials the subject of much
investigation.~\cite{glass;s87,li;prb92,silve;prb95,balza;prb85} \stz\
is an example of a II-VI semiconductor pseudobinary alloy that can be
made over the entire range of compositions.~\cite{ohtan;jac92} II-VI
alloys are becoming increasingly important because they are often used
as the basis for magnetic semiconductors with the additional alloying
of small amounts of Mn on the metal
sublattice.~\cite{fiede;nat99,giebu;jap88} The recent suggestion that
high-speed logical circuits can be made out of devices using spin
diffusion instead of electron diffusion (so-called ``spintronics'') is
adding extra impetus to research into these
materials.~\cite{ball;nat00} Clearly, it is important to be able to
characterize the atomic and electronic structure of these alloys in
detail.

The study of alloys is complicated by the fact that considerable local
atomic strains are present due to the disordering effect of the
alloying.  This means that local bond-lengths can differ from those
inferred from the average (crystallographic) structure by as much as
0.1~\AA.~\cite{mikke;prl82,petko;prl99} This clearly has a significant
effect on calculations of electronic and transport
properties.~\cite{li;prb92} To fully characterize the structure of
these alloys it is necessary to augment crystallography with local
structural measurements.  In the past the extended x-ray absorption
fine structure (XAFS) technique has been extensively
used.~\cite{mikke;prl82,boyce;prb85,boyce;jcg89} More recently the
atomic pair distribution function (PDF) analysis of powder diffraction
data has also been applied to get additional local structural
information from \iga\
alloys.~\cite{petko;prl99,petko;jap00,jeong;prb00;unpub} In that case
high energy x-rays combined with good resolution and a wide range of
momentum transfer allow the In-As and Ga-As nearest neighbor peaks to
be resolved. In this paper we describe PDF measurements of the II-VI
alloy \stz\ from neutron powder diffraction measurements using the new
General Materials Diffractometer (GEM) at ISIS.  In these measurements
the distinct Zn-Se and Zn-Te bonds, which differ in length by just
$\Delta r= 0.14$~\AA , could be distinguished demonstrating the
quality of the data from GEM.

Both ZnTe and ZnSe have the zinc-blende structure
($\mbox{F}\bar{4}3\mbox{m}$) where the Zn atoms and Te, Se atoms
occupy the two interpenetrating face-centered-cubic (fcc) lattices. In
the alloys the lattice parameter of \stz\ interpolates linearly
between the end member values consistent with Vegard's
law.~\cite{vegar;zp21} However, both XAFS
experiments~\cite{boyce;prb85,boyce;jcg89} and
theory~\cite{cai;prb92i,cai;prb92ii,mouss;prb92,schab;prb91} show that
the atomic nearest neighbor (nn) distances deviate strongly from
Vegard's law.  Rather, they stay closer to their natural lengths found
in the end-member compounds: $\mbox{L}^{\mbox{\tiny 0}}_{\mbox{\tiny
Zn-Te}}$=2.643(2)\AA$\mbox{ }$ and $\mbox{L}^{\mbox{\tiny
0}}_{\mbox{\tiny Zn-Se}}$=2.452(2)\AA.

A limitation of the XAFS method for studying the local structure of
alloys is that it only gives information about the first and second
neighbor bond-lengths and information about the bond-length
distributions with less accuracy. In this study the PDF analysis of
neutron powder diffraction data is used. The PDF is the instantaneous
atomic number density-density correlation function which describes the
atomic arrangement of the materials. It is the Sine Fourier transform
of the experimentally observed total scattering structure function,
$S(Q)$, obtained from a powder diffraction experiment. Since the total
scattering structure function includes both the Bragg and diffuse
scattering, the PDF contains both {\it local} and {\it average} atomic
structure yielding accurate information on short and intermediate
length-scales. Previous high resolution PDF studies on \iga\ were
carried out using high energy x-ray
diffraction.~\cite{petko;prl99,jeong;prb00;unpub} This yielded data
over a wide $Q$-range ($Q$ is the magnitude of the scattering vector)
which resulted in the very high real-space resolution required to
separate the nearest neighbor peaks from In-As and Ga-As.  The high
$Q$-range coverage and $Q$-space resolution of the new General
Materials (GEM) Diffractometer at the ISIS neutron source allowed us,
for the first time, to obtain similarly high real space resolution
PDFs of \stz\ using neutrons and to resolve the Zn-Se and Zn-Te bonds
that differ in length by only 0.14~\AA. Furthermore, the data
collection time was only sixty minutes compared to the 12 hours for
the x-ray data with similar quality.  The nn distances and average
peak widths are fit using model independent techniques. The PDFs of
the full alloy series have been calculated using a model based on the
Kirkwood potential giving excellent agreement over a wide range of $r$
with no adjustable parameters.


\section{Experimental}\label{sec;exp}

\subsection{Synthesis and Characterization}

\begin{figure}
\begin{center}
  \includegraphics[angle=270,width=3.2in]{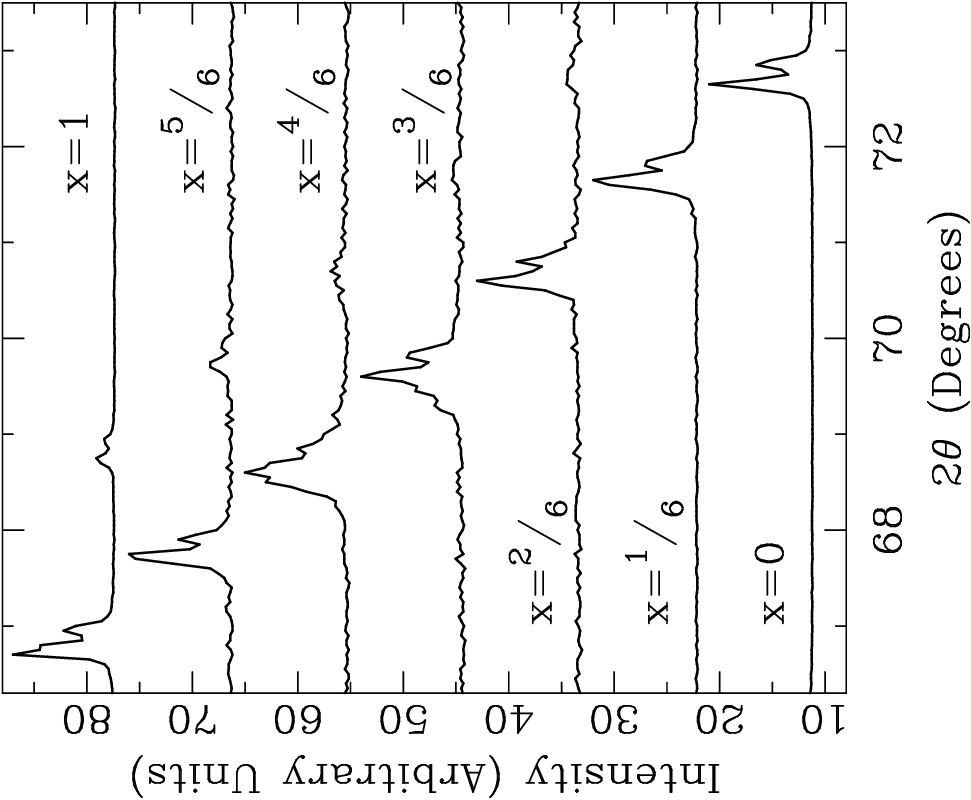} 
  \caption{$\langle 331 \rangle$ and weak $\langle 420 \rangle$ peaks
  of \stz\ measured at 300K using Cu rotating anode.}
  \label{fig;char}
\end{center}
\end{figure}

Finely powdered samples of $\sim$10g of \stz\ were made with
x=$\frac{1}{6}$, $\frac{2}{6}$, $\frac{3}{6}$, $\frac{4}{6}$,
$\frac{5}{6}$.  The starting reagents (zinc selenide, metal basis,
$99.995\%$; zinc telluride, metal basis, $99.999\%$) were finely
ground, mixed in the correct stoichiometry, and sealed in quartz tubes
under vacuum.  The samples were then heated at 900$^\circ$C for 12-16
hours.~\cite{larac;pr57} This procedure (grinding, vacuum sealing, and
heating) was repeated four times to obtain high quality homogeneous
products.  The colors of the solid solutions vary gradually from dark
red (ZnTe) to yellow (ZnSe) as the $x$-value decreases reflecting the
band-gap of the alloy samples smoothly changing in the optical
frequency range.  The homogeneity of the samples was checked using
x-ray diffraction by monitoring the width and line-shape of the
$\langle 400 \rangle$, $\langle 331 \rangle$, $\langle 420 \rangle$,
and $\langle 422 \rangle$ Bragg peaks measured on a rotating anode Cu
K\sb{\alpha} source.  Finely powdered samples were sieved through a
200-mesh sieve then packed into flat plates and measured in symmetric
reflection geometry.  The $\langle 331 \rangle$ and (weak on high
angle side) $\langle 420
\rangle$ peaks are reproduced in Fig.~\ref{fig;char}. 
The double-peaked shape comes from the K\sb{\alpha 1} and K\sb{\alpha
2} components in the beam.  The line-widths are narrow and smoothly
interpolate in position between the positions of the end-members
verifying the homogeneity of the samples.

\subsection{Neutron Measurements and Data Processing}

Time of flight neutron powder diffraction data were measured on
the GEM diffractometer at the ISIS spallation neutron source at
Rutherford Appleton Laboratory in Oxfordshire, UK. The finely powdered
\stz\ samples were sealed inside extruded cylindrical vanadium
containers.  These were mounted on the cold-stage of a helium cryostat
immersed in cold He gas in contact with a liquid He reservoir.  The
temperature of the samples was maintained at 10K using a heater
attached to the cold-stage adjacent to the sample.  The empty
cryostat, an empty container mounted on the cryostat and the empty
instrument were all measured, allowing us to assess and subtract
instrumental backgrounds.  The scattering from a vanadium rod was also
measured to allow the data to be normalized for the incident spectrum
and detector efficiencies.  Standard data corrections were carried out
as described elsewhere~\cite{billi;prb93,wagne;jncs78} using the
program PDFgetN.~\cite{peter;jac00} After being corrected the data are
normalized by the total scattering cross-section of the sample to
yield the total scattering structure function, $S(Q)$. This is then
converted to the PDF, $G(r)$, by a Sine Fourier transform according to
\begin{eqnarray}
G(r)  & = & \frac{2}{\pi}\int_0^\infty Q[S(Q)-1]\sin(Q r) dQ \nonumber\\
      & = & 4 \pi r (\rho(r)-\rho_{\mbox{\tiny 0}}),\label{eq;gr}
\end{eqnarray}
where $\rho(r)$ is the microscopic pair density of the sample and
$\rho_{\mbox{\tiny 0}}$ is the average number density.

\begin{figure}
\begin{center}$\,$
  \includegraphics[angle=270,width=3.2in]{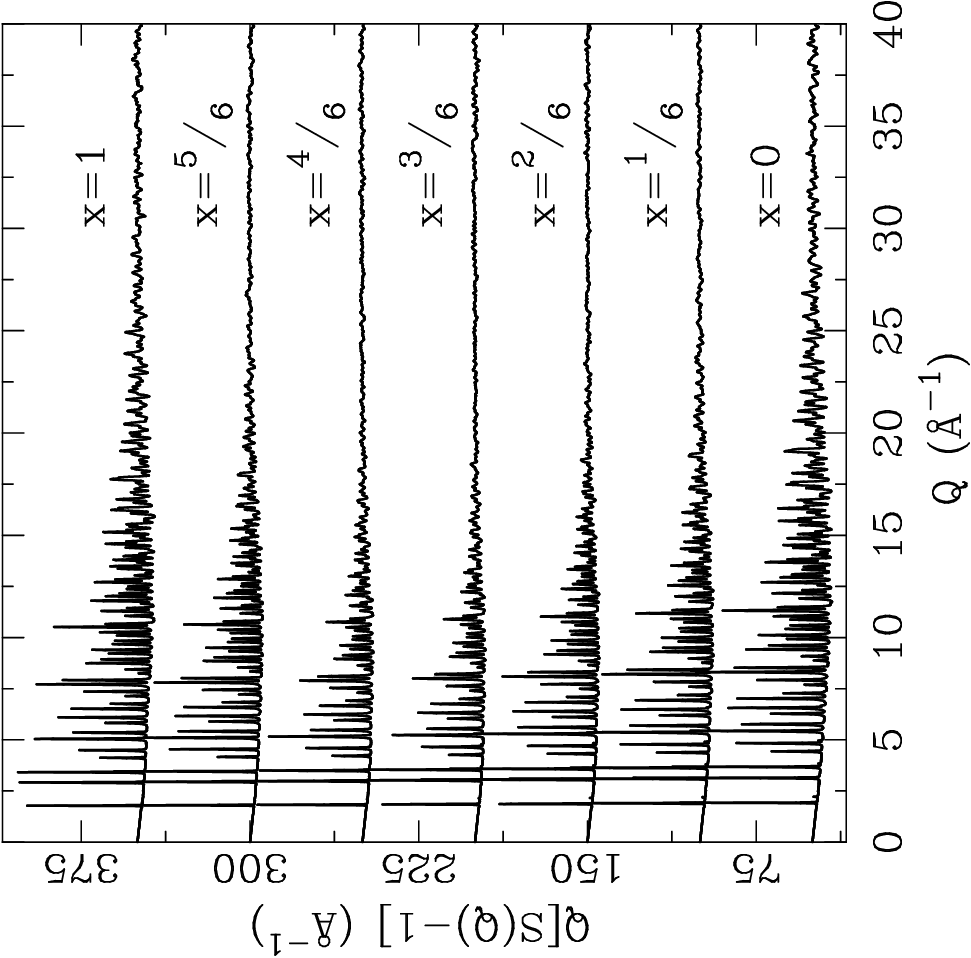}
  \caption{$Q[S(Q)-1]$ for \stz\ measured at 10K.}
  \label{fig;sq_all} 
\end{center}\end{figure}

\begin{figure}
\begin{center}$\,$
  \includegraphics[angle=270,width=3.2in]{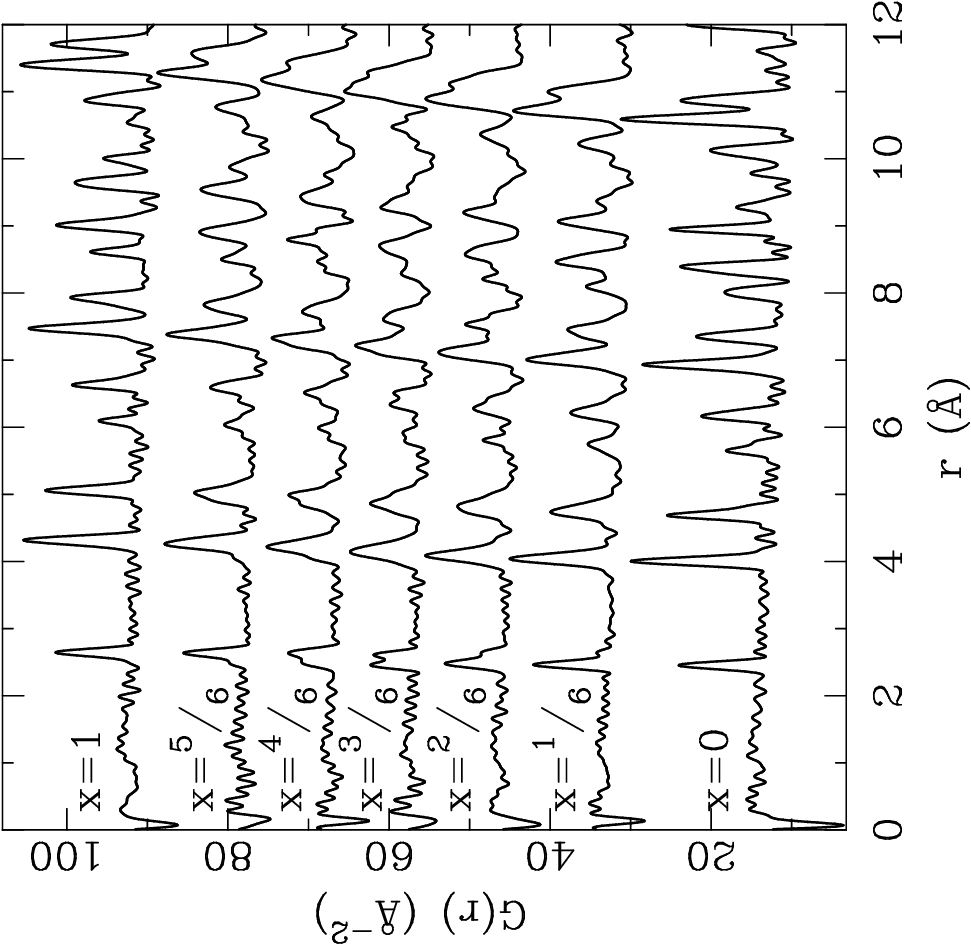}
  \caption{$G(r)=4 \pi r ( \rho(r) - \rho_0 )$ for \stz\ measured at
  10K.}
  \label{fig;gr_all} 
\end{center}\end{figure}

The GEM instrument yields useful diffraction information up to a
maximum $Q$ of greater than 90~\AA$^{-1}$. Unfortunately, due to a
neutron resonance in Te we were forced to terminate the Fourier
transform at a maximum $Q$ of 40~\AA$^{-1}$ in this experiment. This
resulted in nn peaks in these alloys which are resolution limited
rather than sample limited.  This was verified by Fourier transforming
the ZnSe end-member at higher values of $Q$\sb{max}. The nn peak kept
getting sharper up to $Q_{max} = 60$~\AA\sp{-1}.  Nonetheless, the
distinct short and long bond distances are still evident in the alloy
PDFs. At the time of this measurement the backscattering detector
banks on GEM were not operational. With the backscattering banks
yielding better statistics in high Q and adding detector coverage, one
might expect to get similar quality PDFs in a fraction of the
time. The reduced structure functions, $Q[S(Q)-1]$, obtained from the
\stz\ samples are shown in Fig.~\ref{fig;sq_all}
and the resulting PDFs are shown in Fig.~\ref{fig;gr_all}.

\subsection{Method of Modeling}

The PDF can be calculated from a structural model by taking advantage
of the definition of radial distribution function
(RDF),~\cite{billi;prb93} $T(r)$,
\begin{equation}
  T(r) = 4 \pi r^2 \rho(r) = \sum_{i,j} \frac{b_i b_j}{\langle b
  \rangle^2} \delta(r - r_{ij}),
\end{equation}
and substituting the calculated $\rho (r)$ into Eq.~\ref{eq;gr}.  Here
$b_i$ is the scattering length of the $i$th atom, $\langle b \rangle$
is the scattering length averaged over the sample composition and
isotopes, $r_{ij}=|{\bf r}_i-{\bf r}_j|$ is the distance separating
the $i$th and $j$th atoms and the sums are taken over all the atoms in
the entire sample. To approximate the atomic thermal motion we
convolute the delta-functions with Gaussians.  Before being compared
to the data the calculated $G(r)$ is convoluted with a termination
function, $\sin(Q_{max}r)/r$, to account for the effects of the finite
range of the measured data.~\cite{billi;b;lsfd98,thorp;b;lsfd98}

We have used three approaches to obtain structural information from
the PDF.  First, we carry out a model independent analysis by fitting
Gaussian functions to peaks in the RDF.  Next we calculate the PDF
expected from the average crystal structure and refine, using a least
squares approach, atomic displacement (thermal) parameters to obtain
empirically the PDF peak widths in the alloys.  This is done using the
PDF refinement program PDFFIT.~\cite{proff;jac99} Finally, we
calculate the PDF from a Kirkwood potential based model where the atom
positions and thermal broadenings are fully determined by the atomic
potential parameters.

The ZnSe and ZnTe nn distances in the alloys were found by fitting two
Gaussians, convoluted with termination functions to account for the
termination effects, to the nearest neighbor peak in the PDF. Peak
positions and widths were varied.  The relative peak intensities were
constrained to those expected from the alloy composition.  The widths
refined to the same values as the end-members within the errors for
all the alloys.  We thus repeated the fits constraining the peak
widths to have the values refined from the end-member PDFs.  This more
highly constrained fitting procedure resulted in less scatter in the
refined peak positions.

To find the far neighbor peak widths PDFFIT~\cite{proff;jac99} was
used. The zinc-blende crystal structure was used with all the atoms
constrained to lie on their average positions.  Lattice parameters,
scale factor, isotropic thermal factors, and $r$-dependent PDF peak
broadening parameters~\cite{billi;b;lsfd98,proff;jac99} were allowed
to vary but the atoms were not allowed to move off their sites.  In
this way, all of the atomic disorder, static and dynamic, is included
in the refined thermal factors that are giving an empirical measure of
the PDF peak widths at higher-$r$. This approach is better than
fitting unconstrained Gaussians because of the problem of PDF peak
overlap at higher distances.  Clearly this approach does not result in
a good model for the alloy structure but will yield a good fit to the
intermediate PDF in the alloys.  However, it should give reliable
empirical estimates of the width of high-$r$ PDF peaks, even when they
are strongly overlapped, and allows us to separate the disorder on the
cation and anion sublattices.  The PDFs were fit over the range of 3
to 15 \AA$^{-1}$.  This range was selected so the global properties of
the alloys could be fit without influence from the nn behavior.

Potential based modeling to yield realistic alloy structures has been
carried out using a model based on the Kirkwood
potential.~\cite{kirkw;jpc39} This procedure has been described in
detail
elsewhere.~\cite{cai;prb92i,cai;prb92ii,chung;prb97,chung;prb99} The
model consists of 512 atoms arranged in the zinc-blende structure with
periodic boundary conditions where the interatomic force is described
by the Kirkwood potential. The system is then relaxed by moving atoms
to minimize the energy.

The Kirkwood potential can be written as,
\begin{equation}
  V = \frac{\alpha}{2} \sum_{i j} ( L_{i j} - L^0_{i j} )^2 
    +\frac{\beta}{8}L_e^2\sum_{ijk}{\lgroup{\cos\theta_{ijk}+\frac{1}{3}}\rgroup}^2,
\end{equation}
where $L_e^2$ is the nn bond length of an undistorted reference
crystal structure, $L_{ij}$ is the length of the bond between the
atoms $i$ and $j$, and $L_{ij}^0$ is the natural bond-length. In this
definition the bond-stretching, $\alpha$, and bond-bending, $\beta$,
force constants have the same units and
$\theta_{ijk}=\arccos(-\frac{1}{3})$, for an ideal
tetrahedron. Literature values~\cite{cai;prb92} were used for the
bond-stretching and bond-bending parameters obtained from elastic
constant measurements. We also tried optimizing $\alpha$ and $\beta$
by fitting to the ZnSe and ZnTe end-member PDFs; however, the PDFs
calculated using both sets of parameters gave comparable agreement
when compared to the alloy data so we simply report the results
obtained with the literature values of $\alpha$ and $\beta$.  The
values of $\alpha$ and $\beta$ for ZnTe and ZnSe used are shown in
Table~\ref{tab;ab}.
\begin{table}
   \caption{$\alpha$ and $\beta$ reported for \stz\ and \iga.~\protect\cite{cai;prb92ii}}
   \label{tab;ab} 
   \begin{tabular}{crrrr} 
         & $\alpha$~(N/m) & $\beta$~(N/m) & $\beta/\alpha$ & a\sp{**} \\ 
    \hline 
    ZnSe & 33.7 & 4.6 & 0.14 & 0.78 \\ 
    ZnTe & 31.1 & 4.7 & 0.15 & 0.76 \\
    \hline
    InAs & 35.1 & 5.8 & 0.16 & 0.74     \\
    GaAs & 44.3 & 9.2 & 0.21 & 0.70     \\
  \end{tabular}
\end{table}

The PDFs for the alloys are calculated with no adjustable parameters
using the same potential parameters used for the end-members.  The
additional bond-bending parameters present in the alloy due to
Te-Zn-Se type configurations are determined as a geometric mean of the
$\beta$ parameters for the end-members.~\cite{chung;prb97} The thermal
broadening of the PDF is calculated by determining the dynamical
matrix from the potential and projecting out the atomic displacement
amplitudes for each phonon.~\cite{chung;prb97}


\section{Results and Discussion}\label{sec;res}


\subsection{Model Independent Results}\label{sec;res-mod}

\begin{figure}
\begin{center}
  \includegraphics[angle=270,width=3.2in]{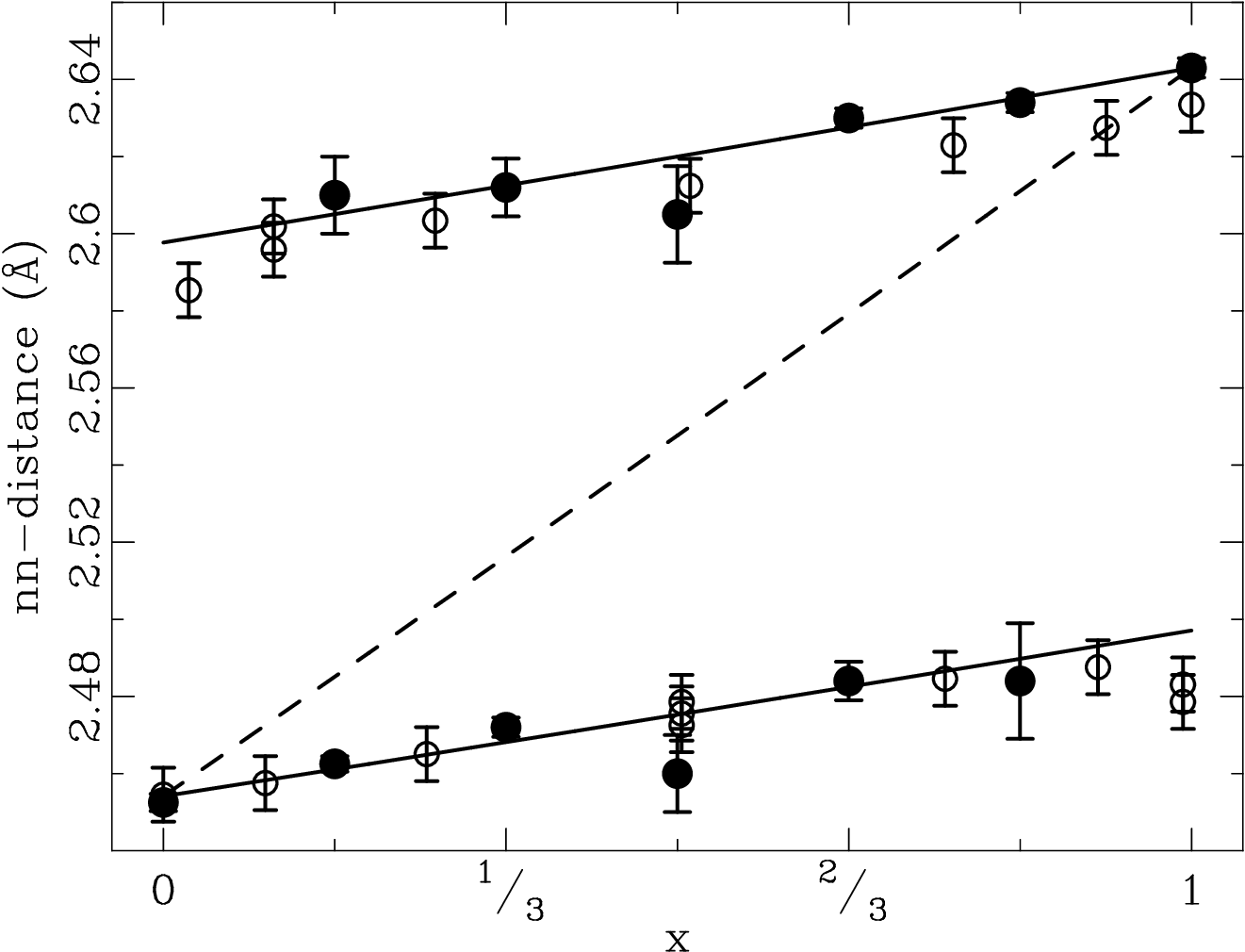} 
  \caption{nn positions from the PDF ($\bullet$), XAFS
  data~\protect\cite{boyce;jcg89} ($\circ$), and Kirkwood model z-plot
  (solid line)~\protect\cite{cai;prb92ii} as a function of
  composition, $x$, for \stz . The dashed line is the average nn
  distance. Note that not all XAFS points had reported error bars so
  they were all set to the same value.}\label{fig;z-plot}
\end{center}
\end{figure}

Upon inspection of the PDFs presented in Fig.~\ref{fig;gr_all} one
will immediately notice the splitting of the first peak. This comes
from the fact that the nn distances of ZnTe and ZnSe stay close to the
end-member values of 2.643(2)\AA\ and 2.452(2)\AA\ respectively. The
positions of each component of the doublet were determined by fitting
Gaussians as described above.  The values for the nn bond-lengths are
shown as filled circles with $2\sigma$ error bars in
Fig.~\ref{fig;z-plot}.
Also plotted in the same Figure as open circles are the nn
bond-lengths determined from an earlier XAFS study by Boyce and
Mikkelson.~\cite{boyce;jcg89} There is clearly excellent agreement
between the two results.  Superimposed on the data are lines which are
the predictions of the Kirkwood model for the nearest neighbor
bond-lengths using the potential parameters given in
Table~\ref{tab;ab}.~\cite{cai;prb92ii} Again, there is excellent
agreement with the data with no adjustable parameters.

In contrast to the local structure, the long-range structure is well
described by the virtual crystal approximation
(VCA).~\cite{nordh;ap31,nordh;ap31b} The VCA assumes that the
structural properties of a crystal alloy all are a linear
interpolation of the end-member values. If this were true in
semiconductor alloys then not only would one be able to find the
lattice parameter of the alloy from Vegard's law but the nn distance
would be, for zinc-blende crystals A$_{1-x}$B$_x$C, $L_{AC}=L_{BC}=
\frac{\sqrt{3}}{4} a$. This is shown in Fig.~\ref{fig;z-plot} as the 
dashed line. 

It is clear that the local bond-lengths remain closer to those in the
end-members than to the prediction of Vegard's law.  In fact the
bond-lengths stay close to the Pauling limit~\cite{pauli;bk67} in
which they would remain completely unchanged in length across the
alloy series.  The deviation from the Pauling limit is attributed to
the disorder in the force constants as we describe below.

\begin{figure}
\begin{center}$\,$
  \includegraphics[angle=270,width=3.2in]{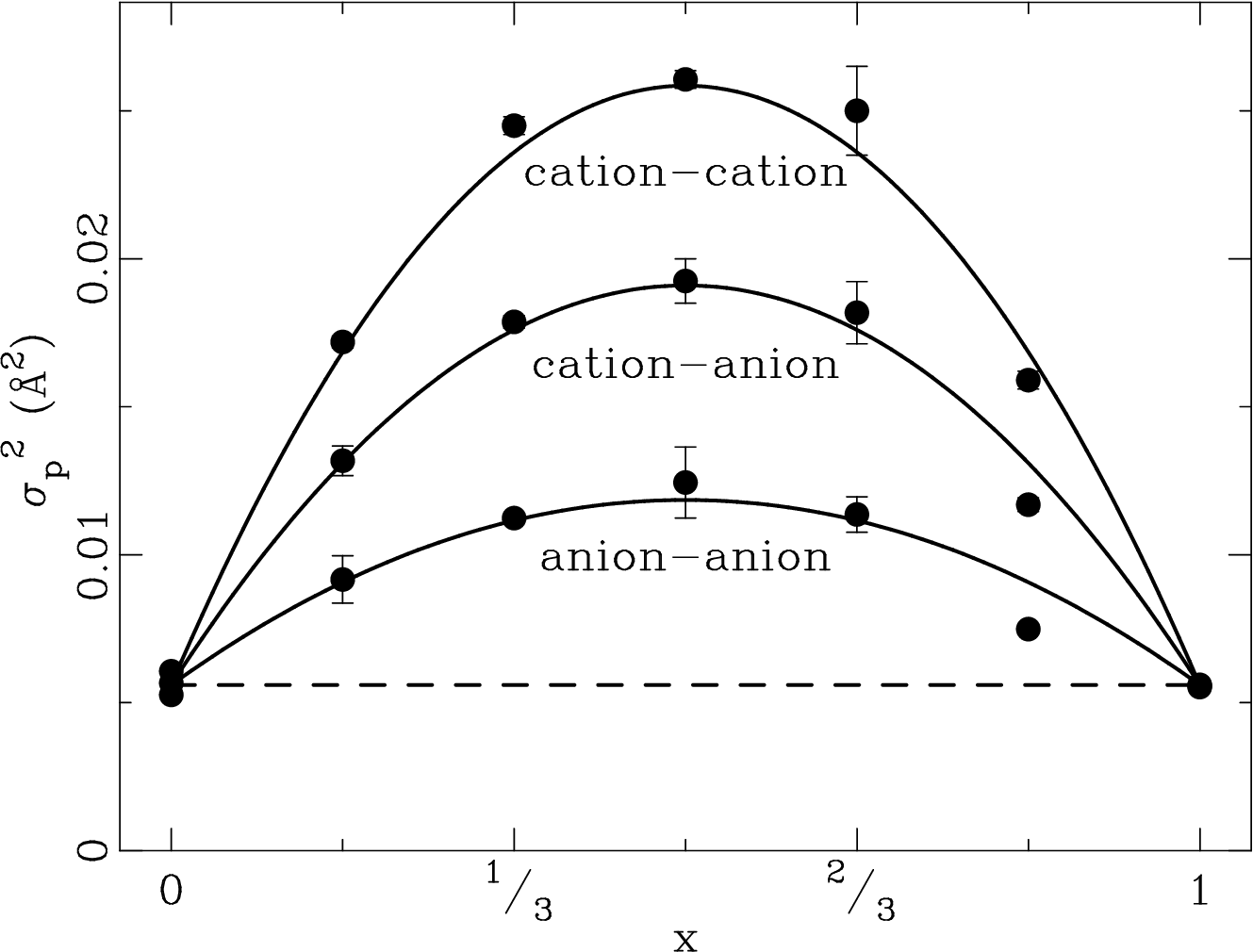}

  \caption{Square of the PDF peak widths for far neighbors as a
  function of composition, $x$. The extracted values (points) are
  plotted with parabolas (lines) to guide the eye. The dotted line is
  at 0.0056\AA$^2$.}
  \label{fig;width} 
\end{center}
\end{figure}

The difference in the local structure of the alloys leads to a large
amount of atomic strain resulting in much broader PDF peaks in the
high-$r$ region of the PDF. In Fig.~\ref{fig;width}
the strain is quantified. The dotted line at
$\sigma_p^2=$0.0056\AA$^2$ represents the mean square width of the PDF
peak, $\sigma_p^2$, attributed to the thermal motion of the atoms
while the parabolas indicate additional peak broadening due entirely
to static strain in the system. The largest peak broadening is seen in
the mean-square width of the Zn-Zn peaks (cation-cation) which is as
much as $5\times$ as large as the mean-square width due to thermal
broadening and zero-point motion at 10~K.  It is also evident that the
disorder is larger on the unalloyed (Zn) sublattice than the mixed
(Se,Te) sublattice similar to the observation in
\iga.~\cite{petko;prl99,jeong;prb00;unpub}


\subsection{Kirkwood Model}\label{sec;res-kir}

The Kirkwood potential~\cite{kirkw;jpc39} is widely used to describe
semiconductor alloys. Petkov {\it et al.}~\cite{petko;prl99} showed
that the Kirkwood model is good at describing \iga, a III-V
semiconductor. \stz\ is a II-VI semiconductor so it is of great
interest to know whether or not the Kirkwood model is equally
successful for this more polar semiconductor alloy.  The values of
$\alpha$ and $\beta$ used in this study are shown in
Table~\ref{tab;ab} with with appropriate values for
\iga\ for comparison.

Other quantities of note are the ratio, $\beta/\alpha$, and the topological
rigidity parameter, \as,~\cite{cai;prb92i} which is a function of
$\beta/\alpha$:
\begin{equation}
\mbox{a}^{**}=\frac{1+1.25(\beta/\alpha)}{1+3.6(\beta/\alpha)+1.17(\beta/\alpha)^2}.
\end{equation}
The topological rigidity parameter, \as, can vary from 0 to 1 and
quantifies the effect of the lattice. The Pauling limit results when
\as=1, the floppy lattice limit. If \as=0 the lattice is rigid and Vegard's
law will hold true locally as well as globally.  As can be seen in
Table~\ref{tab;ab}, the values found for \as\ are close to 0.75 which
appears to be fairly universal for all
semiconductors.~\cite{cai;prb92i}

\begin{figure}
\begin{center}
  \includegraphics[angle=270,width=3.2in]{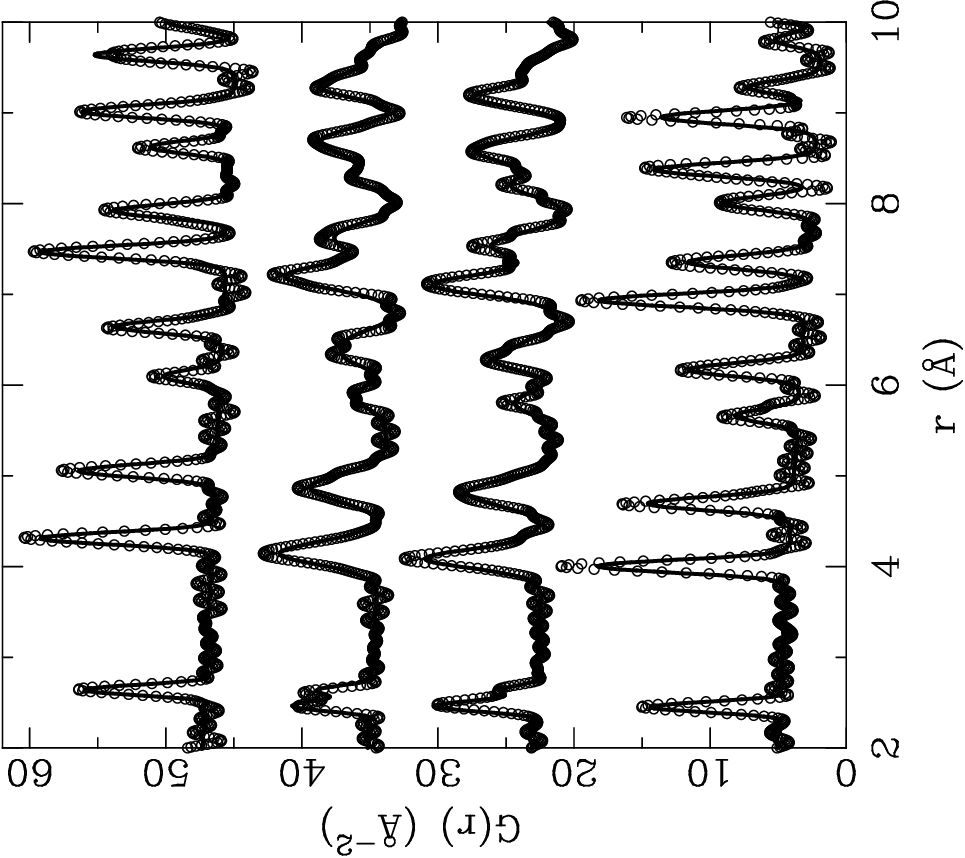}
  \caption{Comparison of the Kirkwood model (lines) and data ($\circ$)
  PDFs for (from top to bottom) ZnTe, ZnSe\sb{3/6}Te\sb{3/6},
  ZnSe\sb{4/6}Te\sb{2/6}, and ZnSe.}  
  \label{fig;kirkwood}
\end{center}
\end{figure}

\begin{figure}
\begin{center}
  \includegraphics[angle=270,width=3.2in]{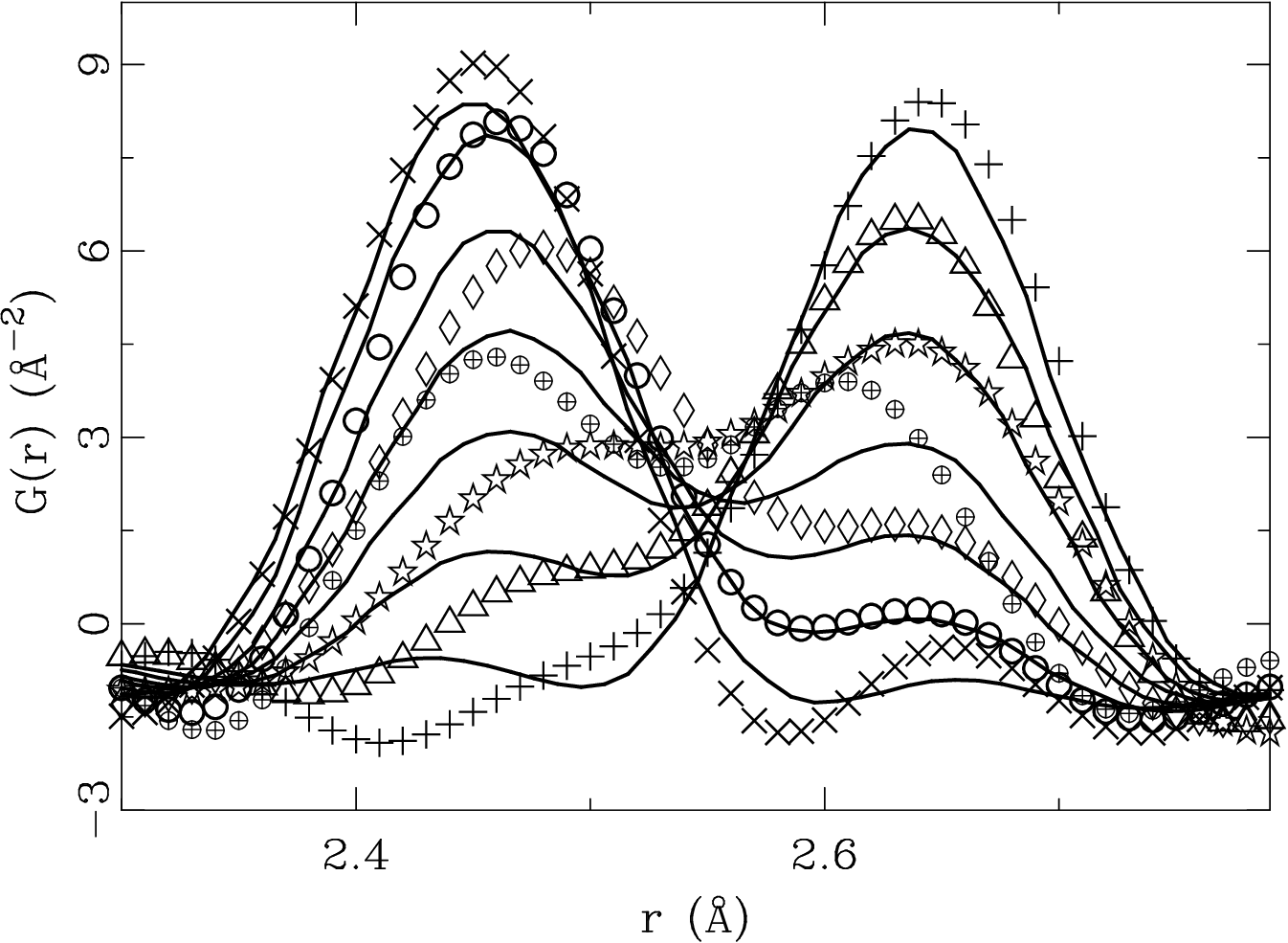}
  \caption{Comparison of the Kirkwood model (lines) and data (points)
  nn distances for \stz\  where $x$ is 0 ($\times$), $\frac{1}{6}$
  ($\circ$), $\frac{2}{6}$ ($\Diamond$), $\frac{3}{6}$ ($\oplus$),
  $\frac{4}{6}$ ($\star$), $\frac{5}{6}$ ($\triangle$), and 1 ($+$).}
  \label{fig;nn-kirkwood}
\end{center}
\end{figure}

In this study, the z-plot shown in Fig.~\ref{fig;z-plot} and PDFs from
each alloy composition were all calculated using the Kirkwood
parameters appropriate for the end-members with no adjustable
parameters. Fig.~\ref{fig;kirkwood}
shows PDFs obtained from the Kirkwood model plotted with measured PDFs
for characteristic compositions. The model is very successful in
matching both the short {\it and} longer-range behavior of the PDFs
for all alloy compositions.  The measured and calculated PDF peaks of
the nearest neighbor bonds are shown on an expanded scale in
Fig.~\ref{fig;nn-kirkwood}.
It is clear that the model based on the Kirkwood potential does a very
satisfactory job of explaining both the PDF peak positions and widths
and appears to produce a very satisfactory model for the structure of
these II-VI alloys.


\subsection{Comparison with \iga}\label{sec;iga}

\begin{figure}
\begin{center}
  \includegraphics[angle=270,width=3.2in]{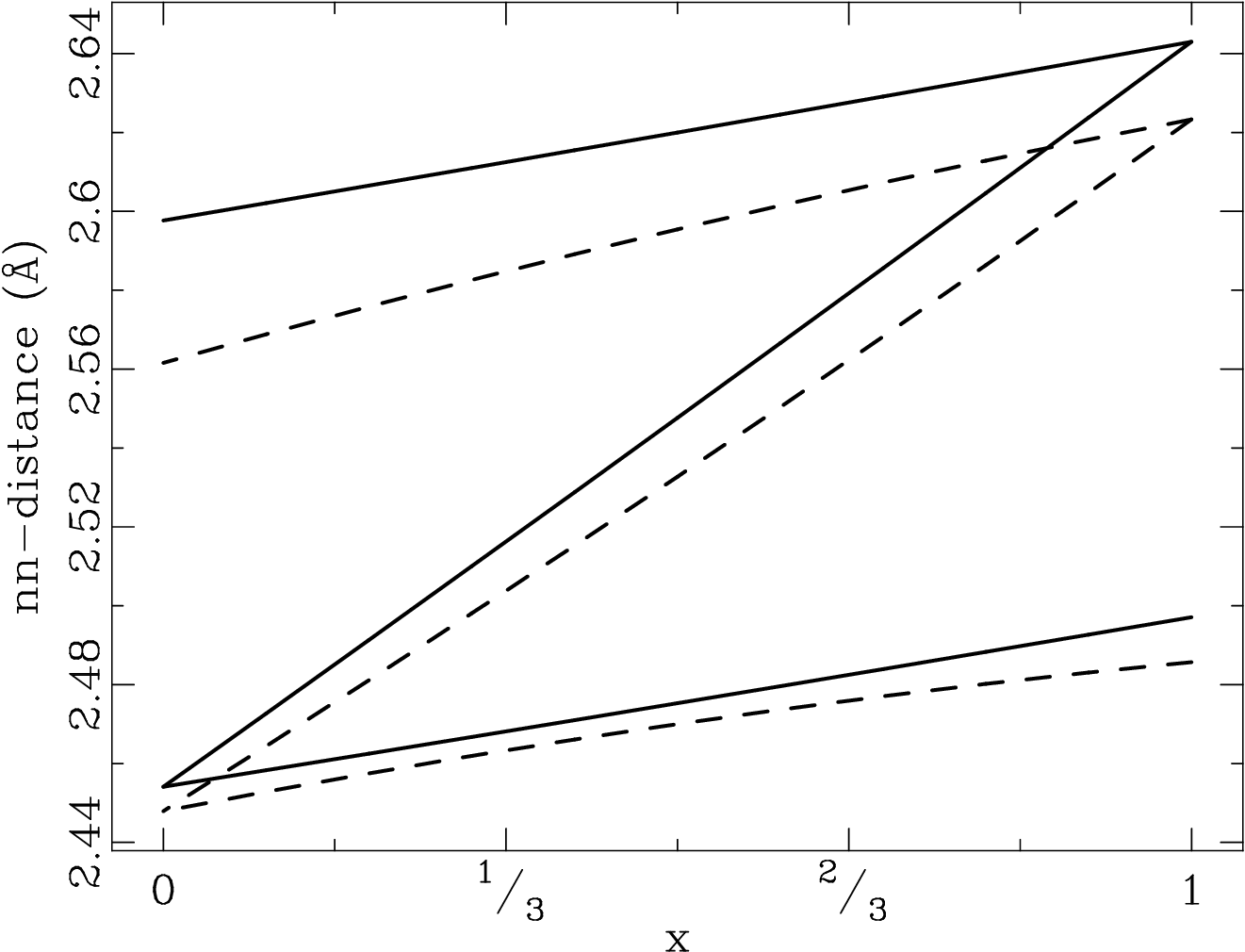}
  \caption{Comparison of the theoretically calculated \stz\ (solid
  line) and \iga\ (dashed line) z-plots.}
  \label{fig;z-compare}
\end{center}
\end{figure}

The results found for \stz\ are not entirely unexpected. In a previous
study of \iga\ similar results were obtained.~\cite{petko;prl99} With
high real space resolution measurements now possible, direct
observation of the nn distances in addition to the static strain in
the system is observed. The basis for the comparison between \stz\ and
\iga\ is that the two systems are both semiconductor alloys with
zinc-blende structures. However, they vary in a couple of important
aspects.  The salient difference is number of valence electrons. \stz\
are more polar alloys and so might be expected to have bonding with
more ionic character than \iga\ which should give rise to smaller
$\beta$ values since ionic bonding is less directional than covalent
bonding. Indeed the bond-bending magnitudes are less in \stz\ than
\iga\ (Tab.~\ref{tab;ab}). However, the nearest neighbor bonds are
stiffer in \iga . This is presumably also due to the lower polarity of
this material resulting in greater orbital overlap and covalency.  The
result is that the $\beta/\alpha$ ratio and \as\ are similar for the
two systems. Since it is this ratio, rather than the values of
$\alpha$ and $\beta$ themselves, that has the greatest impact on the
structure of the alloy, we find very strong similarities in the local
structures of \iga\ and \stz .  For example, the z-plots from both
systems are plotted on the same scale in Fig.~\ref{fig;z-compare}.

The similarity in the z-plots is even more striking because of the
similar bond-lengths of the end-members in these two alloy systems.
The ionic radii for the atoms in these two alloy series are reproduced
in Table~\ref{tab;radii}.
\begin{table}
   \caption{Ionic radii from literature when atoms are in tetrahedral
   covalent bonds.~\protect\cite{kitte;bk96}} 
   \label{tab;radii}
   \begin{tabular}{c|ccc}
   A - B & r\sb{\mbox{\tiny A}} (\AA) & r\sb{\mbox{\tiny B}} (\AA) & r\sb{\mbox{\tiny A+B}} (\AA) \\
   \hline
   GaAs & 1.26 & 1.18 & 2.44 \\
   InAs & 1.44 & 1.18 & 2.62 \\
   ZnSe & 1.31 & 1.14 & 2.45 \\
   ZnTe & 1.31 & 1.32 & 2.63
  \end{tabular}
\end{table}
Despite the ionic radii themselves being somewhat different, it is
clear that the sums of the ionic radii of the end-members yield values
that are within 0.01~\AA\ of each other.

As well as the z-plots of \iga\ and \stz\ matching rather well, the
observed magnitude of the mean-square width of the high-$r$ PDFs peak
are rather similar in the two alloy series.  This can be seen by
comparing Fig.~\ref{fig;width} with Fig.~4 in
Ref.~\onlinecite{petko;prl99}.  For example, the static strain
contribution to the PDF peak widths of the unalloyed site has a
maximum at 0.027\AA\sp{2} and 0.023\AA\sp{2} for $x=0.5$ in \stz\ and
\iga , respectively.

It thus appears that the structure of the alloys is principally
determined by the differences in bond-length of the end-members and by
the $\beta/\alpha$ ratio rather than by the absolute values of
$\alpha$ and $\beta$ or the absolute values of the ionic radii
themselves.  It has been shown~\cite{cai;prb92i,cai;prb92ii} that the
$\beta/\alpha$ ratios of tetrahedral semiconductors are somewhat
universal resulting in \as\ values close to 0.75 for a wide range of
semiconductors.  It also does not appear to matter whether the cation
or anion sublattice is alloyed.  The unalloyed sublattice accommodates
the majority of the atomic scale strain.


\section{Conclusion}\label{sec;con}

From high real space resolution PDFs of \stz\ we conclude the
following. In agreement with earlier XAFS results and the Kirkwood
model the Zn-Se and Zn-Te bond-lengths do not take a compositionally
averaged length but remain close to their natural lengths. Direct
measurement of this was allowed by the new GEM instrument at ISIS. The
bond-length mismatch creates considerable local disorder which
manifests itself as broadening in the PDF peak widths and can be
separated into thermal motion and static strain. \stz\ was compared
with \iga . Despite having different polarity the atomic strains in
both systems are very similar and both are well modeled by the
Kirkwood potential based model.  This suggests that the atomic strains
in tetrahedral semiconductor alloys are quite universal depending
principally on the bond-length mismatch of the end-members and the
ratio of the bond-stretch to bond-bending forces.


\acknowledgements

The authors thank M. F. Thorpe for making available Kirkwood model
programs and for useful discussions concerning the Kirkwood model and
to V. Petkov for useful discussions.  This work was supported by DOE
through Grant No. DE-FG02-97ER45651.
%
%
%

\begin{thebibliography}{10}

\bibitem{glass;s87}
A.~M. Glass,
\newblock Science {\bf 235}, 1003 (1985).

\bibitem{li;prb92}
Z.~Q. Li and W.~Potz,
\newblock Phys. Rev. B {\bf 46}, 2109 (1992).

\bibitem{silve;prb95}
A.~Silverman, A.~Zunger, R.~Kalish, and J.~Adler,
\newblock Phys. Rev. B {\bf 51}, 10795 (1995).

\bibitem{balza;prb85}
A.~Balzarotti, N.~Motta, A.~Kisiel, M.~Zimnal-Starnawska, M.~T. Czy{\. z}yk,
  and M.~Podg{\' o}rny,
\newblock Phys. Rev. B {\bf 31}, 7526 (1985).

\bibitem{ohtan;jac92}
H.~Ohtani, K.~Kojima, K.~Ishida, and T.~Nishizawa,
\newblock J. Alloy Cmpd. {\bf 182}, 103 (1992).

\bibitem{fiede;nat99}
R.~Fiederling, M.~Kelm, G.~Reuscher, W.~Ossau, G.~Schmidt, A.~Wang, and
  L.~Molenkamp,
\newblock Nature {\bf 402}, 787 (1999).

\bibitem{giebu;jap88}
T.~M. Giebultowicz, J.~J. Rhyne, W.~Y. Ching, D.~L. Huber, and J.~Furdyna,
\newblock J. Appl. Phys. {\bf 63}, 3297 (1988).

\bibitem{ball;nat00}
P.~Ball,
\newblock Nature {\bf 404}, 918 (2000).

\bibitem{mikke;prl82}
J.~C. Mikkelson and J.~B. Boyce,
\newblock Phys. Rev. Lett. {\bf 49}, 1412 (1982).

\bibitem{petko;prl99}
V.~Petkov, {I-K. Jeong}, J.~S. Chung, M.~F. Thorpe, S.~Kycia, and S.~J.~L.
  Billinge,
\newblock Phys. Rev. Lett. {\bf 83}, 4089 (1999).

\bibitem{boyce;prb85}
J.~B. Boyce and {J. C. Mikkelsen, Jr.},
\newblock Phys. Rev. B {\bf 31}, 6903 (1985).

\bibitem{boyce;jcg89}
J.~B. Boyce and {J. C. Mikkelsen, Jr.},
\newblock J. Cryst. Growth {\bf 98}, 37 (1989).

\bibitem{petko;jap00}
V.~Petkov, {I-K.~Jeong}, F.~Mohiuddin-Jacobs, {Th. Proffen}, and S.~J.~L.
  Billinge,
\newblock J. Appl. Phys. {\bf 88}, 665 (2000).

\bibitem{jeong;prb00;unpub}
{I.-K. Jeong}, F.~Mohiuddin-Jacobs, V.~Petkov, S.~J.~L. Billinge, and S.~Kycia,
\newblock Phys. Rev. B  (2000),
\newblock submitted. cond-mat/0008079.

\bibitem{vegar;zp21}
L.~Vegard,
\newblock Z. Phys. {\bf 5}, 17 (1921).

\bibitem{cai;prb92i}
Y.~Cai and M.~F. Thorpe,
\newblock Phys. Rev. B {\bf 46}, 15872 (1992).

\bibitem{cai;prb92ii}
Y.~Cai and M.~F. Thorpe,
\newblock Phys. Rev. B {\bf 46}, 15879 (1992).

\bibitem{mouss;prb92}
N.~Mousseau and M.~F. Thorpe,
\newblock Phys. Rev. B {\bf 46}, 15887 (1992).

\bibitem{schab;prb91}
M.~C. Schabel and J.~L. Martins,
\newblock Phys. Rev. B {\bf 43}, 11,873 (1991).

\bibitem{larac;pr57}
S.~Larach, R.~E. Shrader, and C.~F. Stocker,
\newblock Phys. Rev. {\bf 108}, 587 (1957).

\bibitem{billi;prb93}
S.~J.~L. Billinge and T.~Egami,
\newblock Phys. Rev. B {\bf 47}, 14386 (1993).

\bibitem{wagne;jncs78}
C.~N.~J. Wagner,
\newblock J. Non-Crystalline Solids {\bf 31}, 1 (1978).

\bibitem{peter;jac00}
P.~F. Peterson, M.~Gutmann, {Th.~Proffen}, and S.~J.~L. Billinge,
\newblock J. Appl. Crystallogr. {\bf 33}, 1192 (2000).

\bibitem{billi;b;lsfd98}
S.~J.~L. Billinge,
\newblock in {\em Local Structure from Diffraction}, edited by S.~J.~L.
  Billinge and M.~F. Thorpe, page 137, New York, 1998, Plenum.

\bibitem{thorp;b;lsfd98}
M.~F. Thorpe, J.~S. Chung, S.~J.~L. Billinge, and F.~Mohiuddin-Jacobs,
\newblock in {\em Local Structure from Diffraction}, edited by S.~J.~L.
  Billinge and M.~F. Thorpe, page 157, New York, 1998, Plenum.

\bibitem{proff;jac99}
{Th. Proffen} and S.~J.~L. Billinge,
\newblock J. Appl. Crystallogr. {\bf 32}, 572 (1999).

\bibitem{kirkw;jpc39}
J.~G. Kirkwood,
\newblock J. Phys. Chem. {\bf 7}, 506 (1939).

\bibitem{chung;prb97}
J.~S. Chung and M.~F. Thorpe,
\newblock Phys. Rev. B {\bf 55}, 1545 (1997).

\bibitem{chung;prb99}
J.~S. Chung and M.~F. Thorpe,
\newblock Phys. Rev. B {\bf 59}, 4807 (1999).

\bibitem{cai;prb92}
Y.~Cai and M.~F. Thorpe,
\newblock Phys. Rev. B {\bf 46}, 15879 (1992).

\bibitem{nordh;ap31}
L.~Nordheim,
\newblock Ann. Phys. (Leipz.) {\bf 9}, 607 (1931).

\bibitem{nordh;ap31b}
L.~Nordheim,
\newblock Ann. Phys. (Leipz.) {\bf 9}, 641 (1931).

\bibitem{pauli;bk67}
L.~Pauling,
\newblock {\em The Nature of the Chemical Bond},
\newblock Cornell Univ. Press, Ithaca, 1967.

\bibitem{kitte;bk96}
C.~Kittel,
\newblock {\em Introduction to Solid State Physics},
\newblock Wiley, New York, 7th edition, 1996.

\end{thebibliography}
%


\end{document}